\documentclass[review]{elsarticle}

\usepackage{ amsmath,amssymb}
\usepackage{graphicx}
\usepackage{subfigure}
\usepackage{dcolumn}
\usepackage{bm}
\usepackage[hidelinks]{hyperref}
\newcommand{\figref}[1]{\figurename{}~\ref{#1}}

\usepackage{color,soul}
\usepackage{algorithm}
\usepackage[noend]{algpseudocode}
\let\Algorithm\algorithm
\renewcommand\algorithm[1][]{\Algorithm[#1]\setstretch{1}}
\usepackage[outdir=./images/]{epstopdf} 
\usepackage{setspace}
\usepackage{textcomp}
\usepackage{tikz}
\makeatletter
\def\BState{\State\hskip-\ALG@thistlm}
\makeatother
\usepackage{lineno,hyperref}

\RequirePackage[normalem]{ulem} 
\RequirePackage{color}\definecolor{RED}{rgb}{1,0,0}\definecolor{BLUE}{rgb}{0,0,1} 

\modulolinenumbers[5]

\journal{Computational Physics Communications}

\begin{document}

\begin{frontmatter}

\title{RaDiO: an efficient spatiotemporal radiation diagnostic for particle-in-cell codes}
\author[1]{M. Pardal\corref{mycorrespondingauthor}}
 \ead{miguel.j.pardal@ist.utl.pt}
\address[1]{GoLP/Instituto de Plasmas e Fusão Nuclear, Instituto Superior Técnico, Universidade de Lisboa, Lisbon, Portugal}%
\cortext[mycorrespondingauthor]{Corresponding author}

\author[1]{A. Sainte-Marie\fnref{fn1}}

\author[1]{A. Reboul-Salze\fnref{fn2}}

\author[1,2]{R. A. Fonseca}%
\address[2]{DCTI/ISCTE Lisbon University Institute, 1649-026 Lisbon, Portugal}%

\fntext[fn1]{Now at Paris-Saclay University, CEA, CNRS, LIDYL, 91191 Gif-sur-Yvette, France}
\fntext[fn2]{Now at, Max Planck Institute for Gravitational Physics (Albert Einstein Institute) Potsdam Science Park Am Mühlenberg 1 D-14476 Potsdam, Germany}
\author[1]{J. Vieira\corref{mycorrespondingauthor}}
\ead{jorge.vieira@ist.utl.pt}

\begin{abstract}
  This work describes a novel radiation algorithm designed to capture the three-dimensional, space-time resolved electromagnetic field structure emitted by large ensembles of charged particles. 
  The algorithm retains the full set of degrees of freedom that characterize electromagnetic waves by employing the Li\'enard-Wiechert fields to retrieve radiation emission. Emitted electric and magnetic fields are deposited in a virtual detector using a temporal interpolation scheme. This feature is essential to accurately predict field amplitudes and preserve the continuous character of radiation emission, even though particle dynamics is known only in a discrete set of temporal steps. Our algorithm retains and accurately captures, by design, full spatial and temporal coherence effects. We demonstrate that our numerical approach recovers well known theoretical radiated spectra in standard scenarios of radiation emission. We show that the algorithm is computationally efficient by computing the full spatiotemporal radiation features of High Harmonic Generation through a plasma mirror in a Particle-In-Cell (PIC) simulation.
\end{abstract}


\end{frontmatter}


\section{\label{sec:Intro}Introduction}
Radiative processes in plasma are ubiquitous in astrophysics~\cite{Tamburini2011} and in laboratory settings. In plasma acceleration experiments, for example, they are important to the development of compact light sources~\cite{PhysRevE.65.056505}, commonly employed in probing ultra-fast processes. Radiation emission mechanisms in plasma result from collective effects associated with the self-consistent dynamics of a large number of charged particles in the presence of strong electric and magnetic fields. \emph{Ab-initio} numerical models, that can capture the motion of single particles, play an important role in this context, not only to validate theoretical advances, but also to predict radiation emission from experiments and in conditions where analytical models are not available.

Among the different numerical techniques, the Particle-in-Cell (PIC)~\cite{Dawson} scheme provides a standard model to compute the motion of ensembles of charged particles.
In its standard version, the PIC scheme consists in a loop that
iteratively computes electric and magnetic fields by solving a discretized version of the
full set of Maxwell's equations in a grid, and then determines the next positions of
the charged particles according to the relativistic Lorentz force.  PIC codes are thus
capable, by design, to retain most classical radiation emission processes.

The resolution required to capture radiation in the PIC algorithm poses quite stringent limitations on the shortest wavelengths that can be captured directly in a simulation,
given that increasing the grid resolution will lead to a significant increase in the computational load.
Consider a relativistic charged particle, with relativistic factor
$\gamma_p$, undergoing a periodic motion with period T:  The corresponding radiation wavelength,
$\lambda_{rad}$, is proportional to $\lambda_{\mathrm{rad}}\propto c T/\gamma_p^2$
Hence, the spatial resolution required to capture $\lambda_{rad}$ is $\gamma_p^2$ times higher than the resolution needed to describe the particle trajectory. Furthermore, because of the Courant--Friedrichs--Lewy condition, the required temporal resolution is also $\gamma_p^2$ higher than standard. This results in an increase of $\gamma_p^4$  operations
per simulation,
pushing the limits of
current computational capabilities, thereby motivating the development of advanced algorithms to compute radiation emission in PIC codes.

The standard approach to avoid the increased computational load and obtain high-frequency
radiation emission from PIC simulations consists in performing additional radiation
calculations outside the PIC loop using particle trajectory information obtained with the
PIC algorithm. Many simulation codes have been developed over the recent years following
this strategy. The code JRAD~\cite{jrad} receives a set of charged particle trajectories
in order to compute the radiated spectra from the Fourier transform of the
Li\'enard-Wiechert potentials; PIConGPU~\cite{PIConGPU,inproceedings,RadGPU} follows a
similar strategy, but can compute the emitted spectrum as the simulation
progresses; the PIC codes OSIRIS~\cite{Fonseca2002} and EPOCH employ Monte-Carlo approaches to compute the spectrum of radiation from QED processes  at run time (see, e.g. Ref. \cite{PhysRevE.95.023210}). These tools have been successfully used to predict the radiation properties of laboratory plasmas (in plasma based accelerators~\cite{xray_tabletop}), Quantum Electrodynamics~\cite{Martins_2015} and astrophysical plasmas.

However, the spatiotemporal profile of radiation is also important in fields such as astrophysics, where it can reveal the properties of rotating black
holes~\cite{Tamburini2011,1904.07923} for example. It can also play an important role in advanced
 microscopy based on twisted light with helical wavefronts~\cite{OAMSTED}.
Furthermore, this approach also provides a natural description of orbital angular momentum of light.
To address this, we propose a new algorithm that retrieves the spatiotemporal radiation profile instead. This complementary approach includes built-in spatial and temporal coherence effects that are important to describe unexplored features of radiation emission, such as superradiant emission~\cite{10.1038/s41567-020-0995-5}, for example.  Our scheme can be used
whenever the charged particle motion is well resolved, regardless of whether the spatial or temporal resolution is sufficient to resolve the resulting  electromagnetic radiation.

The PIC simulation framework provides a direct and natural application to our present work and we focused on the implementation of this algorithm into the OSIRIS code naming our tool RaDiO, which stands for 
\underline{Ra}diation \underline{Di}agnostic for \underline{O}SIRIS.
 This diagnostic
is composed of two distinct
but equally useful counterparts: one implemented as a post-processing tool that uses
previously generated trajectories to find the radiation that was emitted along
them, and the other implemented as a run-time diagnostic for the PIC code
OSIRIS, that uses the simulation data at each time step to compute the radiation.

This paper is structured as follows. In Section~\ref{sec:Theor}, we describe the
theoretical framework behind radiation emission processes, which lays the groundwork
for the development of
the algorithm. Section~\ref{sec:Impl} describes the implementation of the algorithm in
detail, exploring key aspects like the temporal interpolation scheme. In Section~\ref{sec:bench}, we benchmark our code
against theoretical predictions and the results obtained with  other radiation codes.
Section~\ref{sec:hhg} contains the study of the radiation emitted during the reflection of laser pulses by a plasma mirror. And, finally, Section~\ref{sec:concl} presents the conclusions.


\section{\label{sec:Theor}Spatiotemporal electromagnetic field structure}

\begin{figure}[H]
  \centering
  \includegraphics[width=0.7\textwidth]{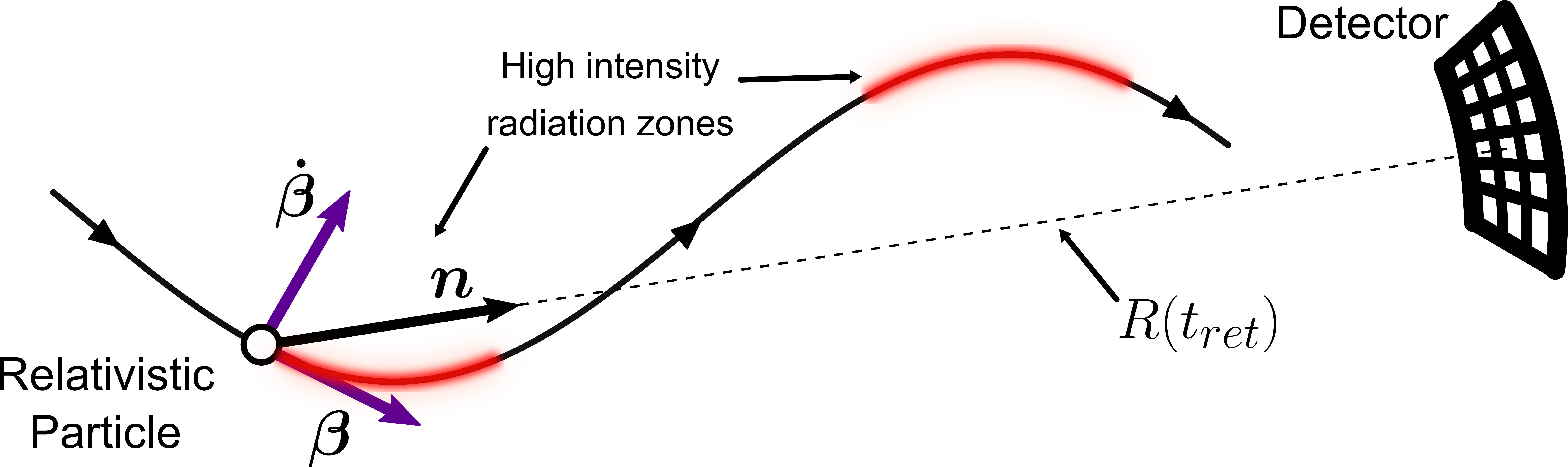}
  \caption{Illustration of the geometry of the radiation emission process and
        relevant quantities. }
  \label{fig:emit}
\end{figure}

The Fourier transformed Li\'{e}nard-Wiechert
fields~\cite{jackson_classical_1999} are commonly employed to predict the radiation spectra from charged particle trajectories. Here, instead, we calculate the Li\'{e}nard-Wiechert fields directly, as these formulas provide the emitted electromagnetic fields at a certain position in space-time. The spatiotemporal Electric~($\mathbf{E}$) and
 Magnetic~($\mathbf{B}$) field structure  of the radiation emitted by
a charged particle according to the Li\'{e}nard-Wiechert formulas is given by:
\begin{equation}
\begin{split}
\mathbf{E}(\mathbf{x},t_{det})&=e\left[\frac{\mathbf{n}-\boldsymbol{\beta}}{\gamma_p^2(1-\boldsymbol{\beta}\cdot\mathbf{n})^3R^2}\right]_{\text{ret}}+\frac{e}{c}\left[\frac{\mathbf{n}\times[(\mathbf{n}-\boldsymbol{\beta})\times\dot{\boldsymbol{\beta}})]}{(1-\boldsymbol{\beta}\cdot\mathbf{n})^3R}\right]_{\text{ret}},\\ \;\;\mathbf{B}(\mathbf{x},t_{det})&=[\mathbf{n}\times\mathbf{E}]_\text{ret},
 \label{eq:lie_wie}
 \end{split}
\end{equation}
\noindent with $\gamma_p=1/\sqrt{1-\beta^2}$. In Equation~(\ref{eq:lie_wie}), the subscript ret
denotes calculations using values at the retarded time, $\mathbf{n}$ is the unit vector oriented
from the particle position to the region in space where we are interested in capturing the emitted
radiation. The virtual region in space-time where radiation is deposited is henceforth denoted as
the \textit{detector} and will be described in more detail in Section~\ref{sec:Impl}. 
In addition, $\boldsymbol{\beta}=\mathbf{v}/c$ and $\dot{\boldsymbol{{\beta}}}= \dot{\mathbf{v}}/c$
are respectively, the particle velocity normalized to  the speed of light, $c$ and the corresponding
acceleration. Here the dot represents the time derivative. The direction of $\boldsymbol{\beta}$
and $\dot{\boldsymbol{\beta}}$ with respect to the virtual detector and $\mathbf{n}$ are 
schematically represented in Figure~\ref{fig:emit}.  Moreover $e$ is the electron charge and the
quantity $R$ is the distance from the particle to the detector. For the purpose of determining 
the radiated fields, the first term in Equation~(\ref{eq:lie_wie}) can be dropped if 
$R\gamma_p^2\dot{\beta}/c\gg 1$. This condition is usually satisfied in the far field 
($R\gg c/\dot{\beta}$) for sufficietly relativistic particles ($\gamma_p\gg1$).
The second term in Equation~\eqref{eq:lie_wie} thus corresponds to emission of propagating electromagnetic waves, describing the so-called acceleration fields.

Equation~\eqref{eq:lie_wie} describes the emitted electric,
$\mathbf{E}$, and magnetic, $\mathbf{B}$, fields at a given position, $x$ and time $t$,
calculated from quantities obtained at the retarded time $t_\text{ret}$. For a given light ray that reaches the detector at a time $t_\text{det}$, $t_\text{ret}$ is the instant of time when emission has occurred. The time of arrival $t_\text{det}$ is given by:
\begin{equation}
t_{det}=t_{ret}+|\mathbf{r}_{part}-R_{cell}\mathbf{n}_{cell}|/c,
\label{eq:tdet}
\end{equation}
\noindent where $\mathbf{r}_{part}$ is the position of the particle and $R_{cell}\mathbf{n}_{cell}$
is the position of the detector's cell. In order to enhance computational performance, it is useful and possible to simplify Equation~(\ref{eq:tdet}) in the far field, which gives ~\cite{jackson_classical_1999}:
\begin{equation}
t_{det}=t_{ret}+R_{cell}/c-\mathbf{r}_{part}\cdot\mathbf{n}_{cell}/c,
\label{eq:ffapprx}
\end{equation}

Supplemented by the additional conditions given by Equations~(\ref{eq:tdet}-\ref{eq:ffapprx}),
Equation~(\ref{eq:lie_wie}) can thus be used to retrieve the full set of spatiotemporal degrees of freedom
of the radiation emitted by accelerated charges. By mapping the emitted radiation at each timestep in
the particle trajectory to the corresponding time of arrival at the detector, the actual temporal
resolution of the relativistic  particle trajectory can be much coarser than the required one to
describe the radiated fields.

\begin{figure}[h!]
      \centering
      \includegraphics[width=0.56\columnwidth]{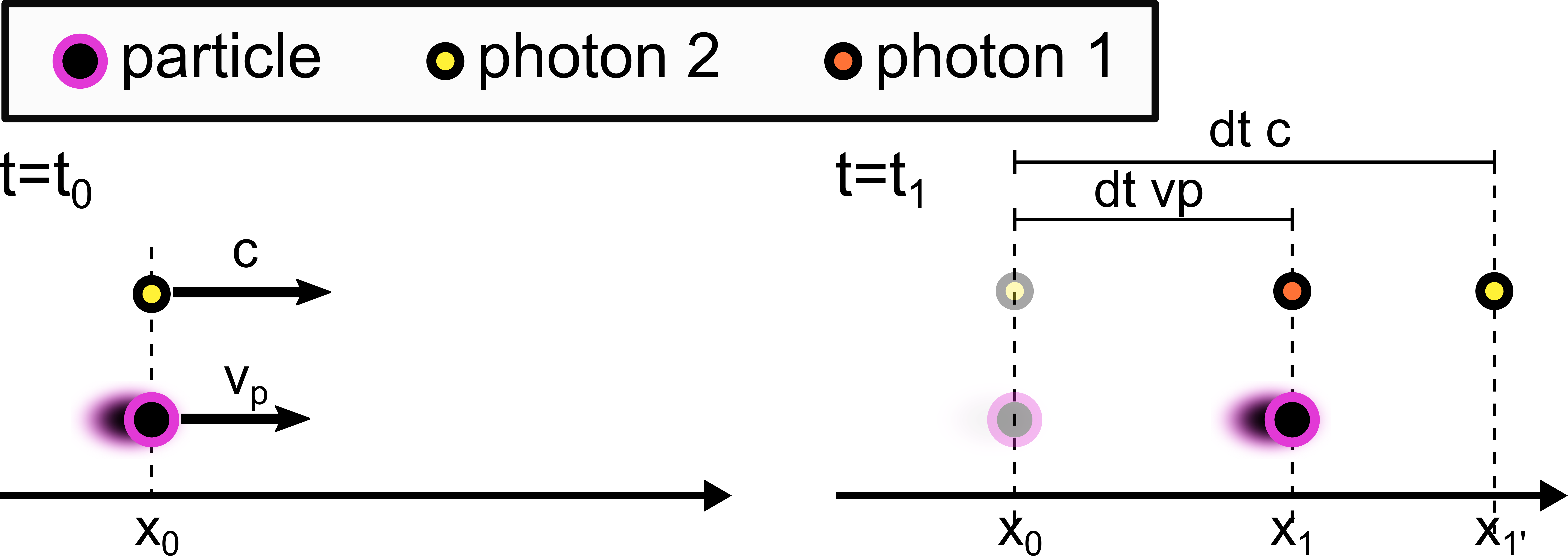}
      \caption{Illustration of radiation emission.  }
      \label{fig:radfreq}
\end{figure}

An estimate of the maximum resolution that can be accurately obtained using
Equations~(\ref{eq:tdet}-\ref{eq:ffapprx})
can be found using the simplified picture shown in \figref{fig:radfreq}: The particle located at $x_0$ emits a photon 1 at $t=t_0$. As the photon travels at c, in the next time step it will have travelled an extra $dt(\text{c}-v_p)$ than the particle, which emits a second photon at $t=t_1$. Considering that a particle emits a photon at every time-step, the time interval between the arrival of two consecutive photons at the detector,
provided that they are emitted by a relativistic particle, is given by Equation~\eqref{eq:timestep}:

\begin{equation}
\label{eq:timestep}
dt_{rad}=dt(1-v_p/c)\simeq dt/\left(2\gamma_p^2\right),
\end{equation}

\noindent with $dt$ being the temporal distance
between emissions and the temporal resolution of the simulation providing for the particle trajectory.

Therefore, we are able to capture radiation with frequencies up to $2\gamma_p^2$ times
larger than the ones used to sample the particle's motion, as our detector time grid can
be as fine as $dt_{\text{det}}=dt/2\gamma_p^2$.
By consequence, the simulation time step can be much larger than the typical period of the emitted
radiation.
It is also important to note that the resolution in the detector should not be increased indefinitely as resolving
time grids finer than $dt/2\gamma_p^2$ could generate non-physical information. 
A thorough analysis of these limits can be found in the Supplementary Material.
The next section describes our implementation of the radiation algorithm and illustrates
the reasons behind the different limits in resolution.

\section{\label{sec:Impl}Algorithm and Implementation}

 The calculations of Equation~(\ref{eq:lie_wie}) can be fully integrated either into a pre-existing code that computes the trajectories of charged particles (e.g. the PIC scheme) or be used as a post-processing tool that
that computes Equation~(\ref{eq:lie_wie}) on a set of pre-calculated trajectories. The algorithm consists of two main parts: calculating and obtaining the radiated fields and depositing them  in a discretized grid. 
In this section we discuss the general steps and approach to incorporate the radiation algorithm considering these two components.

\subsection{Radiation calculation algorithm}

\begin{figure}[h!]
  \centering
  \begin{tikzpicture}
      \node at (0,0) {     \includegraphics[width=0.65\textwidth]{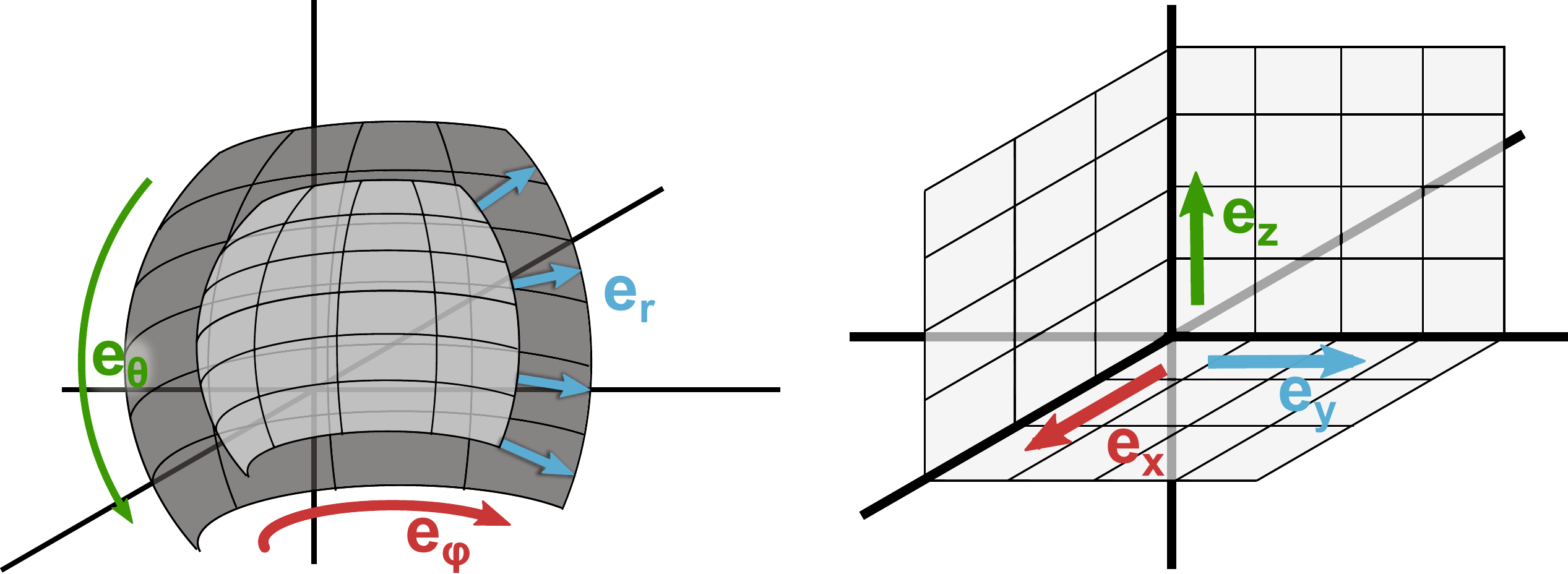}
      };
      \node at (-3.8,1.3) {a)};
      \node at (0.0,1.3) {b)};
      \end{tikzpicture}
      \caption{Spherical (a) and cartesian (b) detectors. The darker spherical grid has a higher radius than the lighter one. All spherical grids are centered in the origin of the coordinate system.}
      \label{fig:exmpldet}
\end{figure}

The virtual detector is a key feature of the radiation diagnostic. It is the
region of space where radiation is tracked during a given time period.
We consider two geometries of the virtual detector,
(i) a spherical one [\figref{fig:exmpldet} a)], where the grid is defined using spherical coordinates
($\mathbf{e_\theta},\mathbf{e_\phi},\mathbf{e_r}$) and (ii) a cartesian one [\figref{fig:exmpldet} b)],
where the grid is defined using cartesian coordinates ($\mathbf{e_x},\mathbf{e_y},\mathbf{e_z}$). RaDiO has the capability to compute the radiation in both types of geometries.

\begin{figure}[h!]
  \centering
      \includegraphics[width=0.7\textwidth]{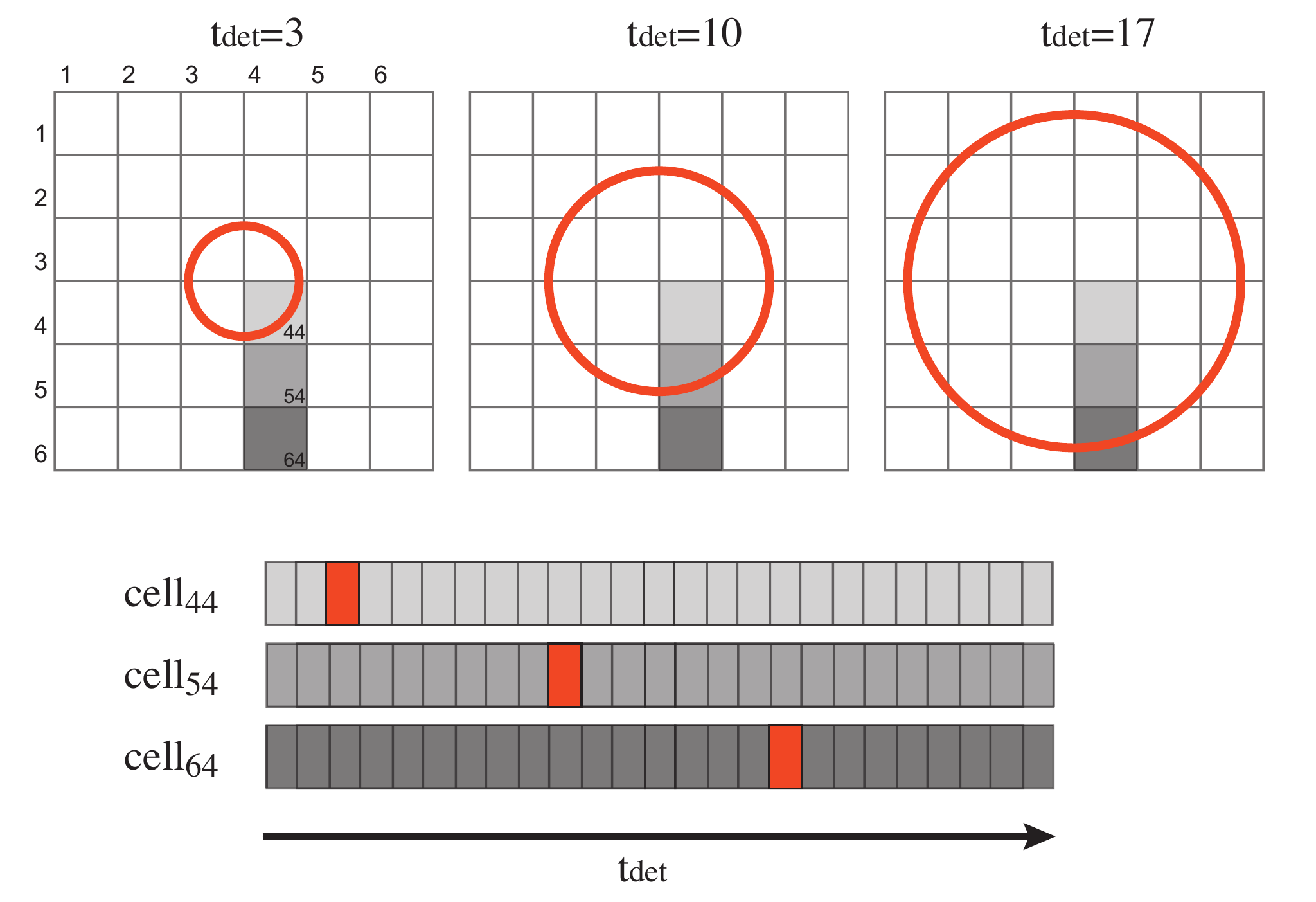}
      \caption{Visual representation of the arrival of radiation emitted by a single particle in a single time step of the simulation at a detector cartesian detector. Top panel: expansion of the intersection between the radiation shell and the detector (in orange). Bottom panel: Time of detection for three distinct cells of the detector.}
      \label{fig:intersect}
\end{figure}

In order to track the emitted radiation at each time step of the trajectory
we need to evaluate Equation~\eqref{eq:lie_wie}
in every cell of the virtual detector. 
The radiation emitted at each time step of a given trajectory lies on a spherical shell that expands from the position of the particle at the time of emission, $t_\text{ret}$, at the speed of light. %
The intersection of the radiation shell with the detector consists of a circumference, whose radius increases with $t_\text{det}$. Figure~\ref{fig:intersect} illustrates this picture, by showing the
intersection of the radiation shell with a cartesian detector. The top of Figure~\ref{fig:intersect}
shows the detector at three different $t_\text{det}$. The bottom of \figref{fig:intersect} shows the
radiation arriving at each one of the highlighted cells as a function of $t_\text{det}$, which can be
calculated using Equation~\eqref{eq:tdet} or Equation~\eqref{eq:ffapprx}.

The illustration of Figure~\ref{fig:intersect} suggests a clear approach to track the radiation reaching the detector from the emission of one particle at a given time step $t_\text{ret}$:
Loop through each spatial cell of the detector and to compute $t_\text{det}$ at which radiation arrives.
All the required quantities to compute Equation~(\ref{eq:lie_wie}) are known or can be easily
calculated (see additional details below). This approach features the quality of avoiding loops through
the temporal cells in the detector. Thus, radiation computing time becomes independent from the temporal resolution of the detector and the total computing time is proportional to the number of time steps in the PIC simulation, $N_\text{t\_PIC}$ multiplied by the number of particles, $N_\text{part}$ multiplied by the number of spatial cells  in the detector, $N_\text{sp\_cell}$.

This approach is summarized in Algorithm~\ref{alg:rad_dep}. It comprises two different
loops: one through the particles that emit radiation (denoted as radiative particles) and another through the detector
spatial cells. The quantities $t,\,R,\,\mathbf{n},\,\mathbf{\beta}$, $\dot{\beta}$ and $t_{det}$
are required in order to evaluate Equation~\eqref{eq:lie_wie}.
All of these quantities are either readily available or can be directly calculated from other quantities that are available in the simulation,
such as the position of the particle ($\mathbf{x_{part}}$), the momentum of the particle ($\mathbf{p}$) and the time of emission $t$, as well as quantities that are
part of the radiation module, such as the position of each detector cell
$\mathbf{x_{cell}}$ or the previous velocity of the particle $\beta_{prev}$.
These calculations are also shown in Algorithm~\ref{alg:rad_dep}.

Because $t_\text{det}$ can be computed at each time of emission, $t_\text{ret}$, using
Equation~(\ref{eq:tdet}) or Equation~(\ref{eq:ffapprx}), it is in principle possible to conceive a temporally
gridless detector.
This approach could provide a very accurate description of the radiated fields, particularly if complemented by a post-interpolation scheme with the goal of retaining the continuous nature of radiation emission. Such approach, however, would require storing as many spatial detector arrays as the number of steps in the particle trajectory, for every  particle in the simulation ($N_\text{t\_PIC}\times N_\text{part}\times N_\text{sp\_cell}$). High memory consumption would thus be the main limitation of such algorithm.
To face this issue, RaDiO 
deposits radiation in a grid detector with up to 3 dimensions (1 temporal dimension and up 
to 2 spatial dimensions) with the spatial cells being distributed according to a spherical 
or cartesian geometry, and uses a temporal interpolation scheme to mimic continuous radiation emission between two consecutive PIC time-steps for every particle in the simulation.

 \begin{algorithm}[H]
\caption{Radiation calculation and depositing}\label{alg:rad_dep}
\begin{algorithmic}[1]
\Procedure{RadiationCalculator}{}
\ForAll{$particle$ \textbf{in} simulation}
\State{$\mathbf{\beta}= \text{velocity}(particle)=\mathbf{p}/\sqrt{|\mathbf{p}|^2+1}$}
\State{$\dot{\mathbf{\beta}}= \text{acceleration}(particle)=(\beta-\beta_{prev})/dt$}
\ForAll{$cell$ \textbf{in} detector}
\State{$\textit{R}= \text{distance}(particle,cell)=|\mathbf{x_{part}}-\mathbf{x_{cell}}|$}
\State{$\mathbf{n}= \text{direction}(particle,cell)=(\mathbf{x_{part}}-\mathbf{x_{cell}})/R$}
\State{$t_{det}=\textit{R}/c+t$}
\State{$t_{det,prev}=\textit{R}_{prev}/c+t-dt$}
\If{$t_{\text{det}}min<t_{\text{det}}<t_{\text{det}}max$}
\State{$\Call{RadiationInterpolator}{\mathbf{E}(\mathbf{n},\mathbf{\beta},\mathbf{\dot{\beta}}),t_{\text{det}},t_{\text{det,prev}}}$}
\EndIf
\EndFor
\EndFor
\EndProcedure
\end{algorithmic}
\end{algorithm}

The implementation shown in Algorithm~\ref{alg:rad_dep} can be applied to both post-processing diagnostics, which calculates the radiation given a set of pre-calculated trajectories, and to run-time diagnostics, in which the radiation calculations are performed at run time during the trajectory calculation. In the latter scenario, the calculation and deposition of the emitted radiation can
take place in a sub-step of the particle push loop, created specifically for that purpose.
This  sub-step comes right after pushing the particles,
in such a way that the newly calculated positions and momenta can be used, in conjunction with the corresponding stored values from the previous iteration, to compute the required quantities to determine the radiated fields.
In the post-processing version, all required quantities can be readily calculated by considering the positions and momenta from consecutive time-steps.

\subsection{Deposition of the radiated fields in a virtual detector}

According to Eqs. (2) and (3), each PIC simulation timestep corresponds to a given detector time. In general, consecutive time steps in the trajectory will deposit radiation in non-consecutive detector time cells. A simple prescription that only deposits the radiated fields in the temporal cells that are closest to the predictions given by Eqs. (2) and (3) will therefore generate noisy radiation patterns that are non-physical because particles emit radiation continuously. To re-gain the continuous character of radiation emission, and remove the artificial noise induced by the discretization of the trajectories in time, RaDiO interpolates the fields emitted by each particle between every two consecutive PIC time steps.

The interpolation scheme in RaDiO assumes that particles radiate constant fields between each consecutive PIC timestep.
In order to deposit the fields across different temporal cells, we weigh the contribution of
each deposition by the time until the next deposition. In fact, the value of the radiation in a time
slot is the integral of the radiation in the interval delimited by two consecutive detector time-steps.
Incidentally, real-life applications often employ an \textit{integrator detector}, which takes the information about radiation arriving in-between detector time steps into account.
This deposition scheme can be implemented by following Algorithm~\ref{alg:t_intp}, below.

\begin{algorithm}[H]
      \caption{Radiation interpolation}\label{alg:t_intp}
      \begin{algorithmic}[1]
      \Procedure{RadiationInterpolator}{}
      \State{$n_{\text{slot}}=\text{slot}(t_\text{array},t_{\text{det}})$}
      \State{$n_{\text{slot,prev}}=\text{slot}(t_\text{array},t_{\text{det,prev}})$}
      \State{$n_{\text{itr}}= t_{\text{slot,prev}}$}
      \State{$t_{\text{tmp}}= t_{\text{det,prev}}$}
      \While{$n_{\text{itr}}<n_{\text{slot}}$}
      \State{$\text{scale\_factor}= (t_{\text{array}}[${{$n_{\text{itr}}$}$+1]-t_{\text{tmp}})/dt_\text{det}$}}
      \State{$\textit{\textbf{E}}(cell,n_{itr})=\mathbf{E}(\mathbf{n},\mathbf{\beta},\mathbf{\dot{\beta}})\cdot \text{scale\_factor}$}
      \State{$n_{\text{itr}}= n_{\text{itr}}+1$}
      \State{$t_{\text{tmp}}= t_{\text{array}}[n_{\text{itr}}]$}
      \EndWhile
      \State{$\text{scale\_factor}= (t_{\text{det}}-t_{\text{array}}[n_{\text{slot}}])/dt_\text{det}$}
      \State{$\textit{\textbf{E}}(cell,n_{itr})=\mathbf{E}(\mathbf{n},\mathbf{\beta},\mathbf{\dot{\beta}})\cdot \text{scale\_factor}$}
      \EndProcedure
      \end{algorithmic}
      \end{algorithm}
      
Each variable in Algorithm~\ref{alg:t_intp} is calculated at each PIC time-step 
and for each particle. Here, \textit{slot(...)} is a function that returns the index of the slot in
the detector's time-array ($t_\text{array}$) where $t_\text{det}$ falls,
$t_{det}$ is the time of the current deposition and $n_\text{slot}$ is the corresponding 
time-slot position in the detector array. In addition, $n_\text{itr}$ is an iterator that runs from $n_\text{slot,prev}$, the detector time slot where particle deposited radiation in the previous PIC time-step, until $n_\text{slot}$. The quantity $t_\text{tmp}$ is an auxiliary variable for the calculation of the time difference between depositions. It runs from, $t_\text{det,prev}$, the time of the previous deposition, to $t[n_{\text{itr}}]$, the time for the actual deposition.

\begin{figure}[h!]
      \centering
      \includegraphics[width=0.42\textwidth]{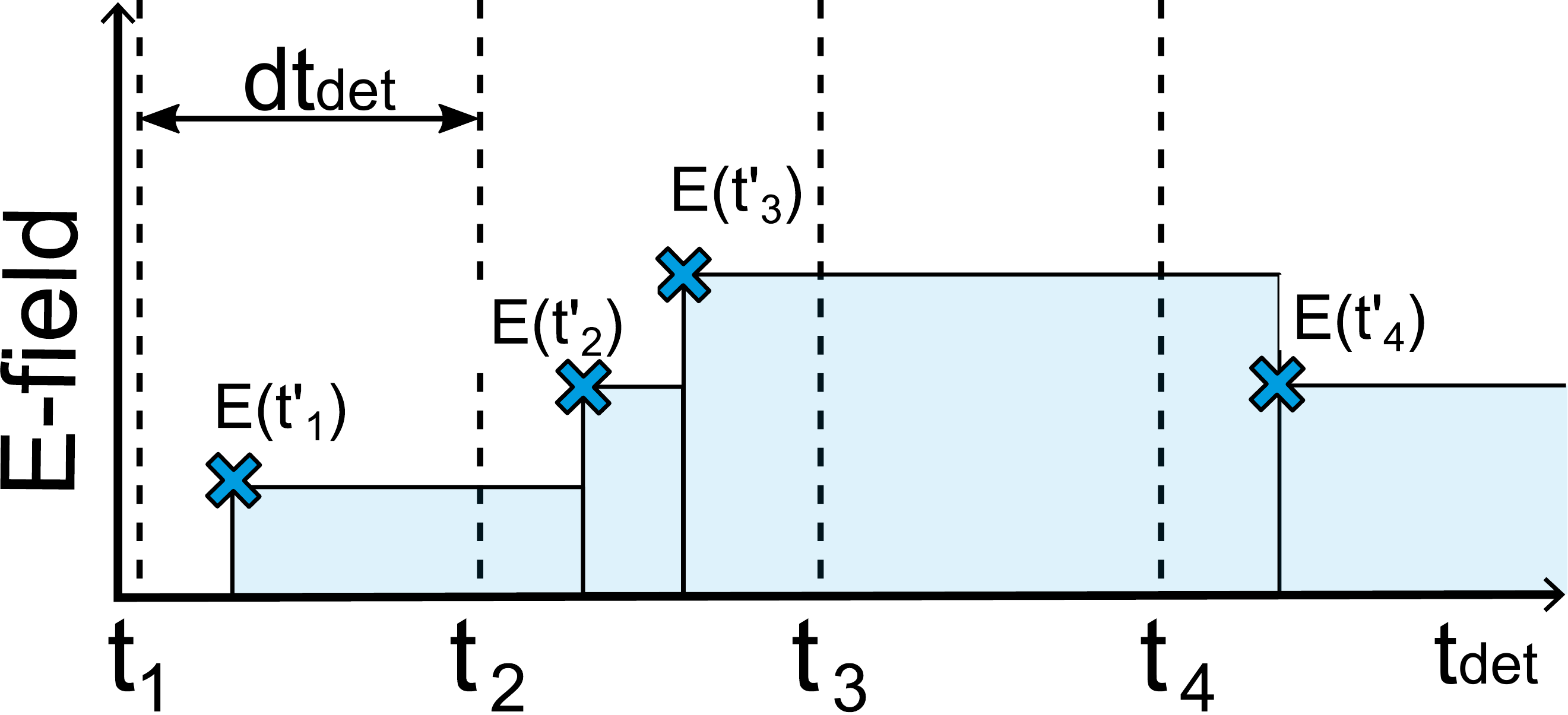}
      \caption{Integrator detector: radiation is scaled by the time until the next deposition. $t_i$ refers to the detector's time grid and $t'_i$ to the
different deposition times.}
      \vspace{-5pt}
      \label{fig:interperf}
\end{figure}

\figref{fig:interperf} shows an example case that clarifies
this deposition scheme.
Each of these depositions correspond to radiation emitted
at a different PIC time step by a single particle. This
interpolation can be performed while the simulation is running, as it only requires
information about the radiated field in the previous time step. In fact, for the
example present in \figref{fig:interperf} the deposition algorithm would go as
follows:
\subitem 1) At PIC iteration 4, radiation arrives at the detector at $t_\text{det}=t'_4$.
\subitem 2) $n_\text{itr}$ is set to 2, the slot of the previous deposition, at $t'_3$,
$t_\text{tmp}$ is set to $t'_3$, the time of the previous deposition, we enter the loop,
the scale factor is calculated: $(t_3-t_\text{tmp})/dt_\text{det}$, with $t_\text{array}[n_{\text{itr}}+1]=t_3$ and
 $E(t'_3)(t_3-t'_3)/dt_\text{det}$ is deposited in the second time slot, $t_2$.
\subitem 3) $n_\text{itr}$ is incremented to 3, $t_\text{tmp}$ is set to $t_3$, the time of the previous deposition, the scale factor is calculated: $(t_4-t_3)/dt_\text{det}$, with $t_\text{array}[n_{\text{itr}}+1]=t_4$ and $E(t'_3)(t_4-t_3)/dt_\text{det}$ is deposited in the third time slot, $t_3$.
\subitem 4) $n_\text{itr}$ is incremented to 4, $t_\text{tmp}$ is set to $t_4$, we exit the loop,
the scale factor is calculated: $(t_\text{det}-t_4)/dt_\text{det}$ and $E(t'_3)(t'_4-t_4)/dt_\text{det}$ is deposited in the  time slot $t_4$.

Using this approach, radiation can be computed and deposited using only the
information from the current and the previous time steps.
This algorithm interpolates radiation coming from a single particle, but can be
repeated for all particles in the simulation, as stated in Algorithm~\ref{alg:rad_dep},
in order to capture radiation from all particles.

\subsection{Practical example: Helical trajectory}

\begin{figure}[h!]
\centering
      \includegraphics[width=0.7\textwidth]{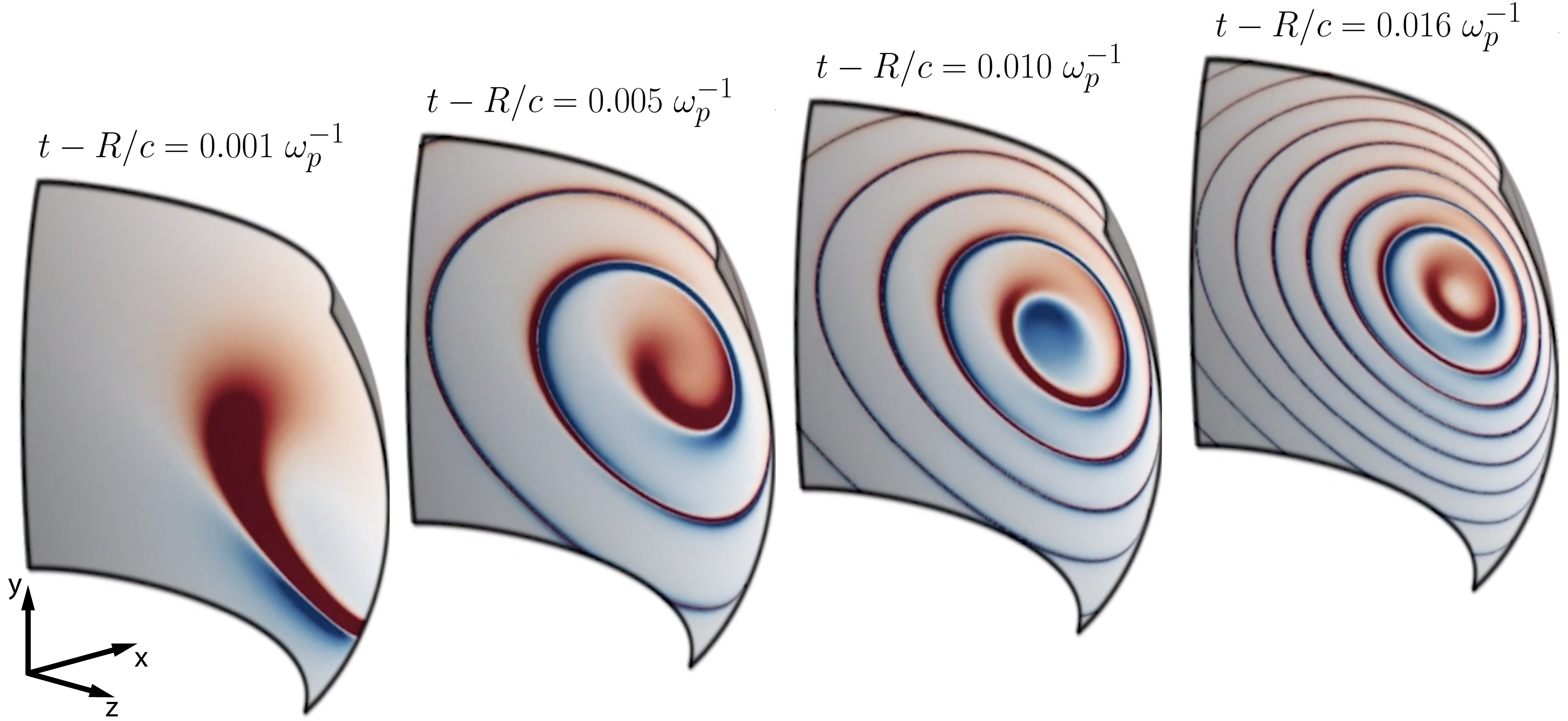}
      \caption{Spatiotemporal signature of the radiation emitted by a particle undergoing a helical trajectory.}
      \label{fig:sphertim}
\end{figure}

Here we look at a practical example, in which an electron with $\gamma_p=57.3$
undergoes an  helical motion with amplitude $0.014\,\text{c}/\omega_p$ and
frequency $\omega_0=\omega_p$, corresponding to a $K$ parameter of $K=0.8$, $K$ is a
trajectory parameter that can be taken as a scaled pitch angle the maximum angle of the particle
 trajectory, normalized to the Lorentz factor $\gamma_p$ and given by $K=\gamma_p r_0 \omega_0/c$.
The helical motion was described by the PIC algorithm
with a temporal resolution of $0.1\;\omega_p^{-1}$. Here, $\omega_p$, is an arbitrary normalizing frequency. The radiation generated by a particle
undergoing such trajectory has a distinctive, expanding spiral spatiotemporal
signature. This is shown in \figref{fig:sphertim}, which represents the radiated
electric field along the $y$ direction deposited onto a spherical 2D detector with an
angular aperture of $0.1$~rad placed in the direction
of the longitudinal motion of the particles ($x$ axis), with radius $R=10^6\;\text{c}/\omega_p$. The temporal resolution of the detector was
$1.33\times 10^{-5}\;\omega_p^{-2}$ and the spatial resolution was $58\;\text{\textmu}\text{rad}$.
\figref{fig:sphertim} shows a snapshot of the detector at four different temporal positions.
 The starting point of the spiral follows the circular motion of
the particle in the $y-z$ plane.
Between each snapshot the radiation spiral makes two turns, thus
the temporal distance between each snapshot is approximately equal to two periods of
the emitted radiation. Given the trajectory parameters, the radiation period was
expected to be of about $\sim 2\times10^{-3}\;\omega_p^{-1}$, about 10 times smaller
than the smaller period that could be resolved using only the PIC algorithm.

\section{\label{sec:bench}Benchmarking}
In order to benchmark our algorithm, we consider the example of a
 single relativistic particle emitting synchrotron radiation.
 Synchrotrons have a magnetic field structure that imposes a sinusoidal trajectory to
relativistic electrons that go through the device, thus leading to the
emission of high frequency photon beams in the X-UV or X-ray regions
of the spectrum.  The trajectory of the particle would then be given by:

\begin{align}
y(t)&=r_\beta\cos{(\omega_\beta t)} \label{eq:traj1} \\
x(t)&=\beta_{x0}\left[t-\frac{r_\beta^2}{8\gamma_{x0}}\left(t-\frac{\cos(2\omega_\beta t)}{2}\right)\right]
\label{eq:traj}
\end{align}

\noindent where $\beta_{x0}$ is the initial velocity of the particle along the longitudinal
$x$ direction, ${\gamma_{x0}=(1-\beta_{x0}^2)^{-1/2}}$ its longitudinal Lorentz factor, $r_\beta$ the amplitude of the sinusoidal trajectory, and $\omega_\beta$ its frequency.

As far as we are aware, the only explicit analytical formulas capturing the spatiotemporal
radiation profile of synchrotron radiation are found in \cite{doi:10.1119/1.1986445}, which
gives a semi-analytical model for the emitted field lines.
However, direct quantitative comparisons between the visual depiction of field lines
and the actual value of the emitted field in a region of space can be difficult
(see Supplementary Material for a qualitative comparison).
On the other hand, the spectral  properties
of radiation are well documented~\cite{PhysRevLett.93.135004,PhysRevSTAB.13.020702}, so we Fourier transformed the data in the virtual detector with respect to time 
and compared these spectra to the theoretical predictions.
The corresponding intensity spectrum  ($I$) with respect to the frequency $\omega$ and solid angle $\Omega$ of the
emitted radiation , valid for ultra relativistic particles as an asymptotic limit expression  ($\gamma_p\gg1$) and assuming very large number of periods in the trajectory, ~\cite{PhysRevE.65.056505}, is given by:

\begin{equation}
\frac{d^2I}{d\omega d\Omega}=\frac{e^2\omega^2\gamma^2}{3\pi^2c\omega_\beta K}\left(\frac{1}{\gamma_p^2}+\theta^2\right)^2\left[\frac{\theta^2}{\gamma_p^{-2}+\theta^2}K^2_{2/3}(\Upsilon)+K^2_{1/3}(\Upsilon)\right], \label{eq:lie_wie_I4}
\end{equation}

\noindent where, $\theta$ is  the observation angle in the direction
perpendicular to the trajectories plane. In addition, $K_n$ is  the modified
Bessel function and $\Upsilon$ is a numerical parameter given by
$\Upsilon=\frac{\omega\gamma_p}{3\omega_\beta K}\left(\gamma_p^{-2}+\theta^2\right)^{-3/2}$,
with $K$ being the aforementioned $K$ parameter.

Equation~\eqref{eq:lie_wie_I4} can be integrated over all angles, returning the
frequency spectrum~\cite{PhysRevE.65.056505}:

\begin{equation}
\frac{dI}{d\omega}=\sqrt{3}\frac{e^2\gamma_p\omega}{c\omega_c}\int_{\omega/\omega_c}^{\infty}K_{5/3}(x)dx,\; \omega_c=\frac{3}{2}K\gamma_p^2\omega_\beta 
\label{eq:sinangint}
\end{equation}

\begin{figure}[h!]
\centering
      \includegraphics[width=0.6\textwidth]{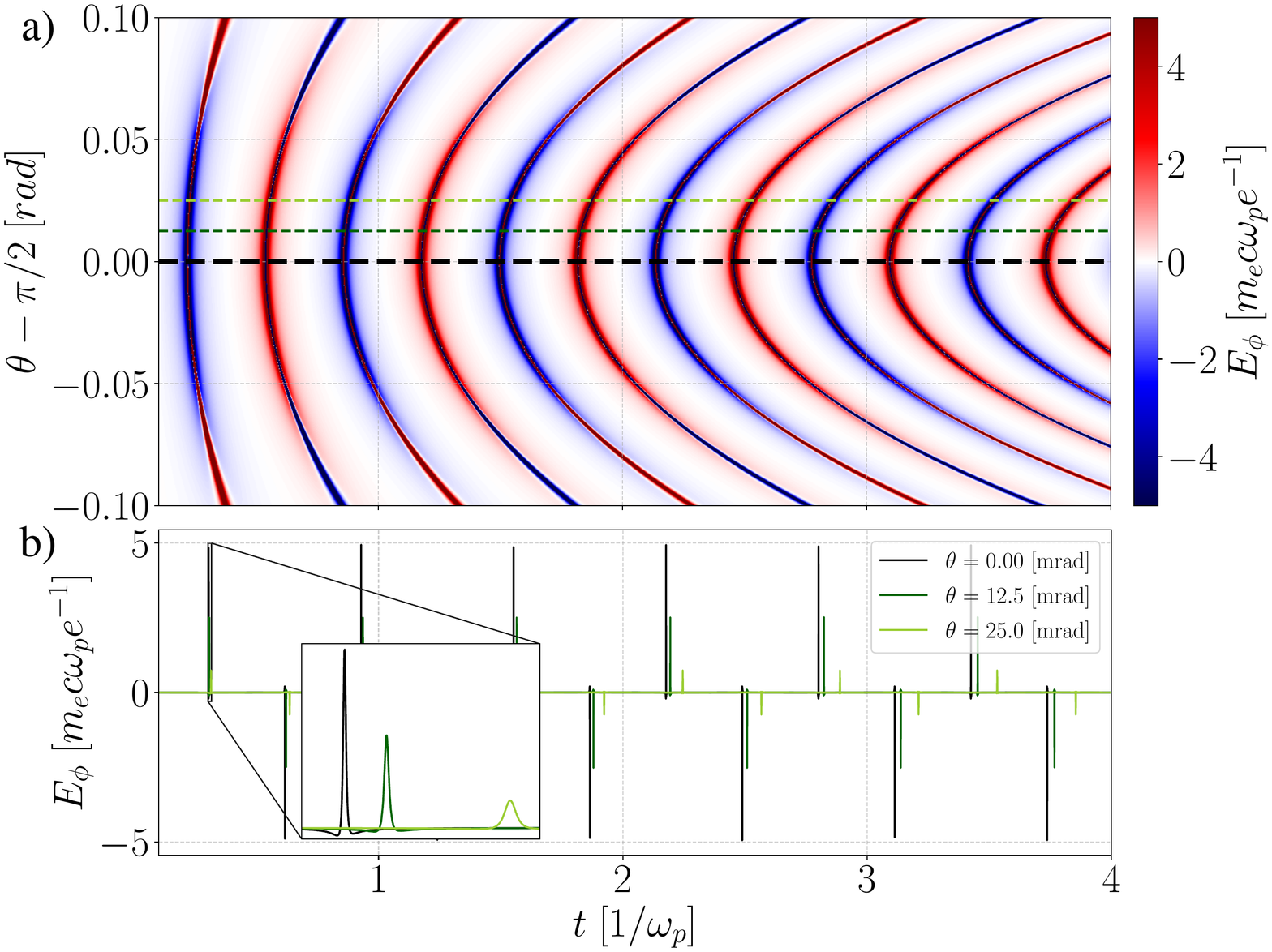}
      \caption{Spatiotemporal signature of the radiation emitted by a particle undergoing a sinusoidal motion in a transverse  detector (a). The lineouts are  shown on the bottom plot (b). Peaks located at smaller $t$ arrive earlier at the detector.}
      \label{fig:sinsign}
\end{figure}

We have benchmarked our algorithm against Equations~\eqref{eq:lie_wie_I4}
and~\eqref{eq:sinangint}.
The benchmarks were performed using the two dimensional sinusoidal trajectory of a
relativistic electron ($\gamma_p=50$) with an amplitude of $r_\beta=2\,c/\omega_p$,
$k_\beta=0.1\,\omega_p/c$ ($K=10$) in the transverse $x-y$ plane
and $dt=0.01c/\omega_p$, where $\omega_p$ is a normalizing frequency.
We simulated a line of a spherical detector, placed
 in the  $z-x$ plane,
$10^5\,c/\omega_p$ away from the axis origin
with an angular aperture of $0.1$\,rad around the $x$ axis. This detector had 512
spatial cells and 131072 temporal cells, resulting in a temporal detector resolution of
$2.98\times10^{-5}c/\omega_p$
The results are shown in \figref{fig:sinsign} which features a plot of the detected
electric field in the $\mathbf{e_\phi}$ direction (perpendicular to the motion plane)
for each spatiotemporal  cell.

The radiation is  composed  of several periodically spaced peaks, whose shape can
be observed in the lineout [\figref{fig:sinsign} b)]. The short burst nature of the radiation (equivalent to a broad band spectrum), consistent with the large
value of the K parameter, is clear from Figure~\ref{fig:sinsign}.
Instead of displaying a purely sinusoidal
profile with a single wavelength, the electric field consists of sharply peaked bursts
containing many different wavelengths. Moreover, it is possible to observe that
consecutive peaks have opposite sign. This is a direct result of the sinusoidal nature of the
electron trajectory in which the acceleration $\dot{\boldsymbol{\beta}}$ switches sign between
peaks. Furthermore, it is possible to note that for
higher angles the radiation bursts arrive later, creating the parabola-like structures that can
be  seen in the upper plot. This delay becomes more significant as the particle
approaches the detector's surface, resulting in a decrease of the curves' aperture.

This result can also be understood in terms of
the spatiotemporal reasoning regarding the estimation for the typical radiation
frequency presented in the previous section. However, instead
of depicting the emitted radiation parallel to the motion of the particle, we picture
them emitted at an angle $\theta$. The temporal distance between the emission and
arrival of light ray emitted at a given longitudinal position $x$ is then given by:
$c\Delta t_{rad}=\sqrt{R^2+x^2-2xR\cos{\theta}}$,
where all quantities are defined as in Equation \eqref{eq:ffapprx}.
This expression %
shows that the time of arrival increases with $\theta$ and also that it is scaled by the longitudinal
position $x$.
Thus, as the particle approaches the detector and $x$ grows larger,
the parabolic structures left on the detector become tighter.

Figure~\ref{fig:sinsign}~b), which depicts lineouts of $E_{\phi},$ also shows that the peaks become wider and less intense for larger angles.
This is in concordance with the predictions for the spectrum [see Equation~\eqref{eq:lie_wie_I4}], which features a
decrease in the number of harmonics for larger angles, resulting in broader and
less intense peaks off-axis.

\begin{figure*}[htp]
\centering
  \includegraphics[width=0.9\textwidth,trim={0 0.1cm 0 0.1cm},clip]{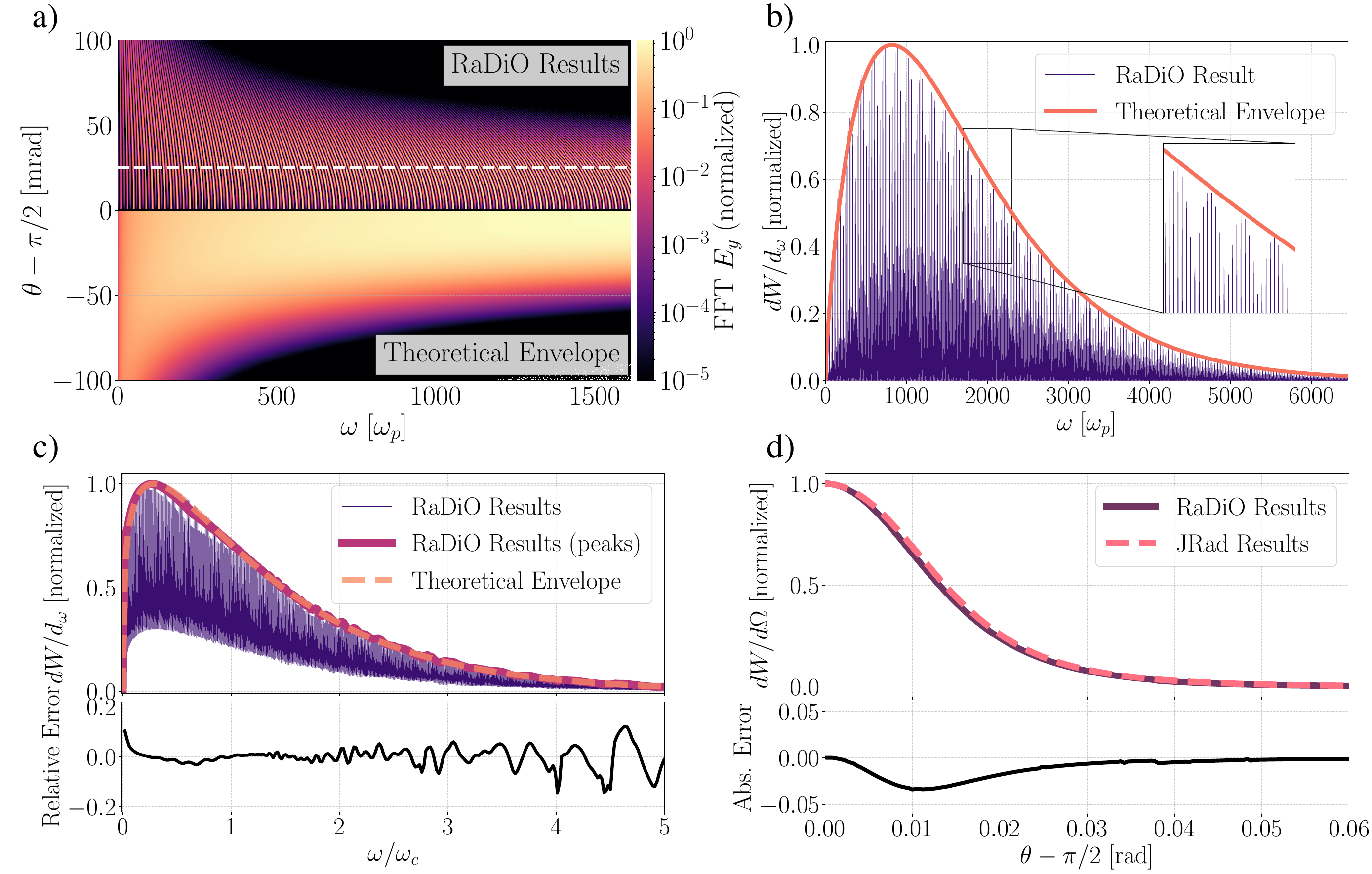}
  \caption{(a) Comparison between the theoretical and simulated  spectra. (b) Comparison between a lineout at $\Delta\theta=0.02$ from both spectra. (c) Angle integrated spectra, both spectra are normalized to 1. The relative error ($I_{\text{RaDiO}}/I_{\text{theor}}-1$) is shown on the inset. (d) Frequency integrated spectra, both spectra are normalized to 1. The absolute error ($I_{\text{RaDiO}}-I_{\text{JRad}}$) is shown on the inset.}
  \label{fig:sincomp}
\end{figure*}

In order to further understand the angular dependent frequency spectra, \figref{fig:sincomp} compares the theoretical result, given by Equation~\eqref{eq:lie_wie_I4}, with the simulated result, given by the Fourier transform over time of the field shown in Figure~\ref{fig:sinsign}~a).
The
spectrum is symmetric with respect to $\theta=0$. Thus, the upper half of \figref{fig:sincomp}~a) ($\theta>0$) shows the simulated results and the bottom half ($\theta<0$) the theory.
As expected, the theoretical line, being the assymptotic limit of a continuous harmonic distribution with a very large number of oscilations in the trajectory~\cite{PhysRevE.65.056505}, corresponds to the envelope of the numerical result, showing excellent agreement. This is evident from the lineout of the radiated spectra displayed in  \figref{fig:sincomp}~b).

The simulated integrated spectrum over all angles, which yields the
frequency distribution of the emitted radiation, can be benchmarked against
Equation \eqref{eq:sinangint}.
Figure 8 shows excellent agreement between numerical and theoretical results, as
the intensity of most peaks matches
the  expected result with small relative error which rises as frequency increases.

To further confirm the validity of our numerical approach, we benchmarked the frequency
integrated spectrum, $dI/d\Omega$ against the spectrum provided by the
post-processing spectral code JRad~\cite{jrad}, which computes the radiated fields using the spectral version of the Liénard-Wiechert potentials. The results of this comparison, shown in \figref{fig:sincomp}~d), are in excellent agreement.

\subsection{Coherence tests}

Because RaDiO captures the emitted fields in space and in time it can also naturally describe temporal and spacial interference effects. This feature is essential to  accurately portrait temporal and spatial coherence, present in superradiant emission scenarios for example. 
This is an intrinsic feature of our spatiotemporal approach, which allows us
to directly obtain the fields radiated by every simulation particle, including interference effects by design.

To test our ability to acurately model temporal and spatial coherence, we ran simulations using two particles with opposite charges and sinusoidal
trajectories, similar the one defined in Equations~\eqref{eq:traj1} and \eqref{eq:traj}.
The two particles, with particle 1 being positively charged and particle 2 being negatively charged underwent this sinusoidal trajectory in perpendicular planes
(particle 1 in plane $x-y$ and particle 2 in plane $x-z$) and the detector was the
same as the one used in the previous section.

\begin{figure*}[htp]
  \includegraphics[width=\textwidth]{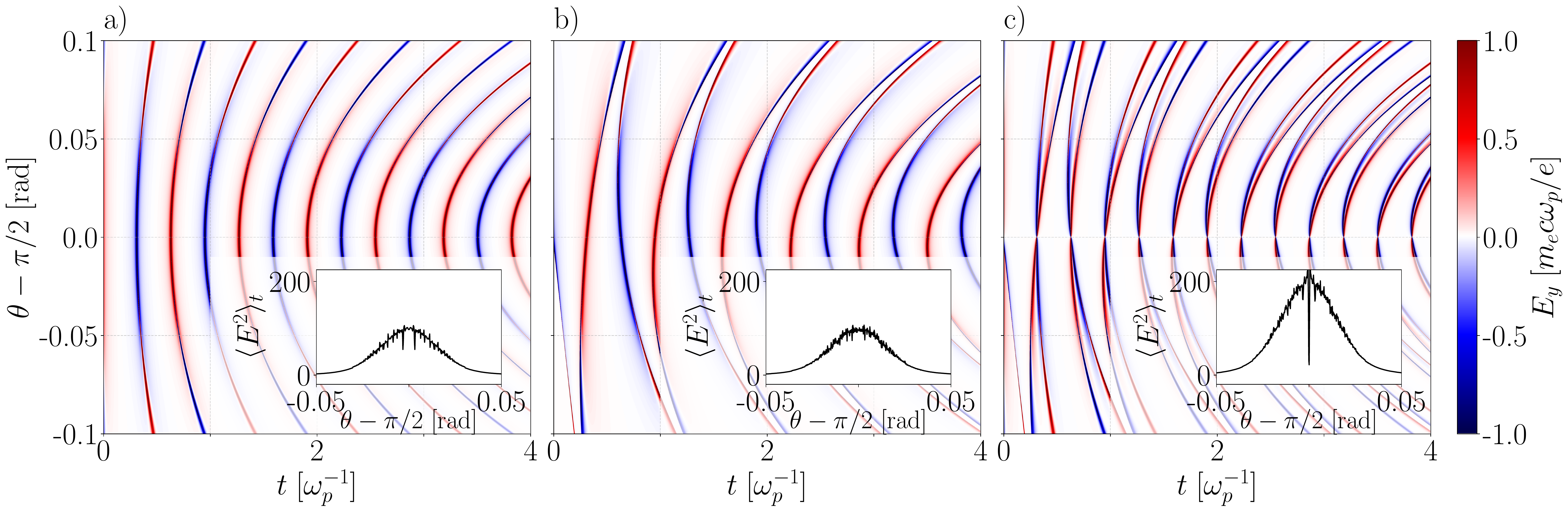}
  \caption{(a) Spatiotemporal profile of the radiation coming from particle 1 (trajectory perpendicular to the detector). (b) Spatiotemporal profile of the radiation coming from particle 2 (trajectory parallel to the detector). (c) Spatiotemporal profile of the radiation coming from both particles. The insets contain the time averaged squared field, $\langle E^2 \rangle_t$.}
  \label{fig:coheranl}
\end{figure*}

\figref{fig:coheranl} shows the simulated radiated electric field profile as a function of $\theta$ and $t_{\mathrm{det}}$ for three different
configurations: one with only particle 1 [\figref{fig:coheranl}~(a)], one with only
particle 2 [\figref{fig:coheranl}~(b)] and other with both particles
[\figref{fig:coheranl}~(c)]. As the two trajectories lie in different planes, the
spatiotemporal signatures of the radiation emitted by each particle are noticeably
 distinct because the detector
plane lies on the plane of the trajectory of particle  2, then being
perpendicular to the plane of particle 1. By comparing \figref{fig:coheranl}~(a) with \figref{fig:coheranl}~(b), we can hence readily identify the radiation coming from each particle in \figref{fig:coheranl}~(c).

As both particles have opposite charges, the field on axis for a
given particle will have the opposite sign as the on-axis field for the other
particle. Thus, the radiation emitted by both particles will interfere
destructively on-axis. This happens exactly at $\theta=\pi/2$. Thus, if we look at the time averaged squared field
(insets in each panel of \figref{fig:coheranl}), we see that although
$\langle E^2 \rangle_t$ is maximum at $\theta=\pi/2$ for the simulations with only
one of the particles (insets of \figref{fig:coheranl}~[a] and \figref{fig:coheranl}~[b]),
the opposite happens when we capture the fields radiated by both particles
(inset of \figref{fig:coheranl}~[ c]).

Our algorithm captures coherence effects  of the simulation particles by default,
but in a PIC code, each particle in the simulation represents a cloud of $N$ real particles 
with a size close to cell size that follow the same dynamics, this is the so called macroparticle aprroximation.
In our code, however, we calculate the radiation emitted by the macroparticles 
in the simulation as if they were point charges with charge equal to the total charge inside the macroparticle ($Nq$). This is in fact 
equivalent to assuming that each of the $N$ particles inside the macroparticle radiates 
coherently. The assumption that they all radiate coherently holds either for all wavelengths if $N=1$, or for wavelengths
larger than the cell size if $N\gg1$. For wavelengths shorter than the cell size, in general, we cannot say it holds, as such an assumption depends 
on information about particles that are not being simulated. 
For example, if standard macroparticle approximation is still valid at scales smaller than the cell size, the emitted radiation
should be incoherent for wavelengths shorter than the cell size and the result should be 
corrected with a filter function (see Supplementary Material for a deeper analyisis).

The detailed study of the conditions that allow assuming that each of the $N$ particles
 inside the macroparticle radiates coherently is out of the scope of this work.
It will be up to the user to decide whether it holds or not. If this assumption does not hold, 
then results given by our code  will be correct for wavelengths larger than the cell size, 
but could be overestimated for wavelengths smaller than the cell size. Nevertheless our 
code can, in general, accurately predict the qualitative aspects of the emitted radiation 
for all wavelengths.

\section{\label{sec:hhg}Example: Radiation from a plasma mirror}

When an electromagnetic wave collides with a target such as a metallic surface or an overdense plasma,
it is unable to propagate and gets reflected. The process of reflection has long been well understood 
and thoroughly explained at the macroscopic level by Maxwell's laws and classical electrodynamics. In the plasma, 
the phenomenon is commonly explored using a fluid theory approach. Such description predicts the damping of the 
wave near the surface of the reflective material (it becomes an evanescent wave) and the appearance of a reflected wave.
At the electron level, however, the phenomenon is not always trivial, in particular at relativistic laser
intensities (with peak normalized vector potential $a_0=eA_0/(m_ec)=1$ ), which lead to High Harmonic Generation (HHG) in plasma mirrors~\cite{Thaury2008,Vincenti2014}.
Several theoretical frameworks have been proposed to describe the underlying mechanisms of HHG, each with different 
regimes of applicability~(see e.g. \cite{HHG_experimental,lichters-pukhov})

PIC simulations are commonly employed to deepen the understanding of the physical processes underlying laser reflection
and harmonic generation in plasma mirrors. An accurate description of HHG in standard PIC simulations, for instance,
is computationally challenging because spatial and temporal PIC grids need to properly resolve the high harmonics.
Thus, to accurately capture high harmonics up to the $10^{th}$ or $100^{th}$ order, PIC simulations require 
spatiotemporal resolution up to one-two orders of magnitude higher than one required to resolve the fundamental harmonic.
The use of RaDiO may thus be computationally advantageous in HHG simulations, as it allows capturing high frequency harmonics
 without increasing the PIC resolution.

\begin{figure}[H]
  \centering
  \includegraphics[width=0.7\textwidth]{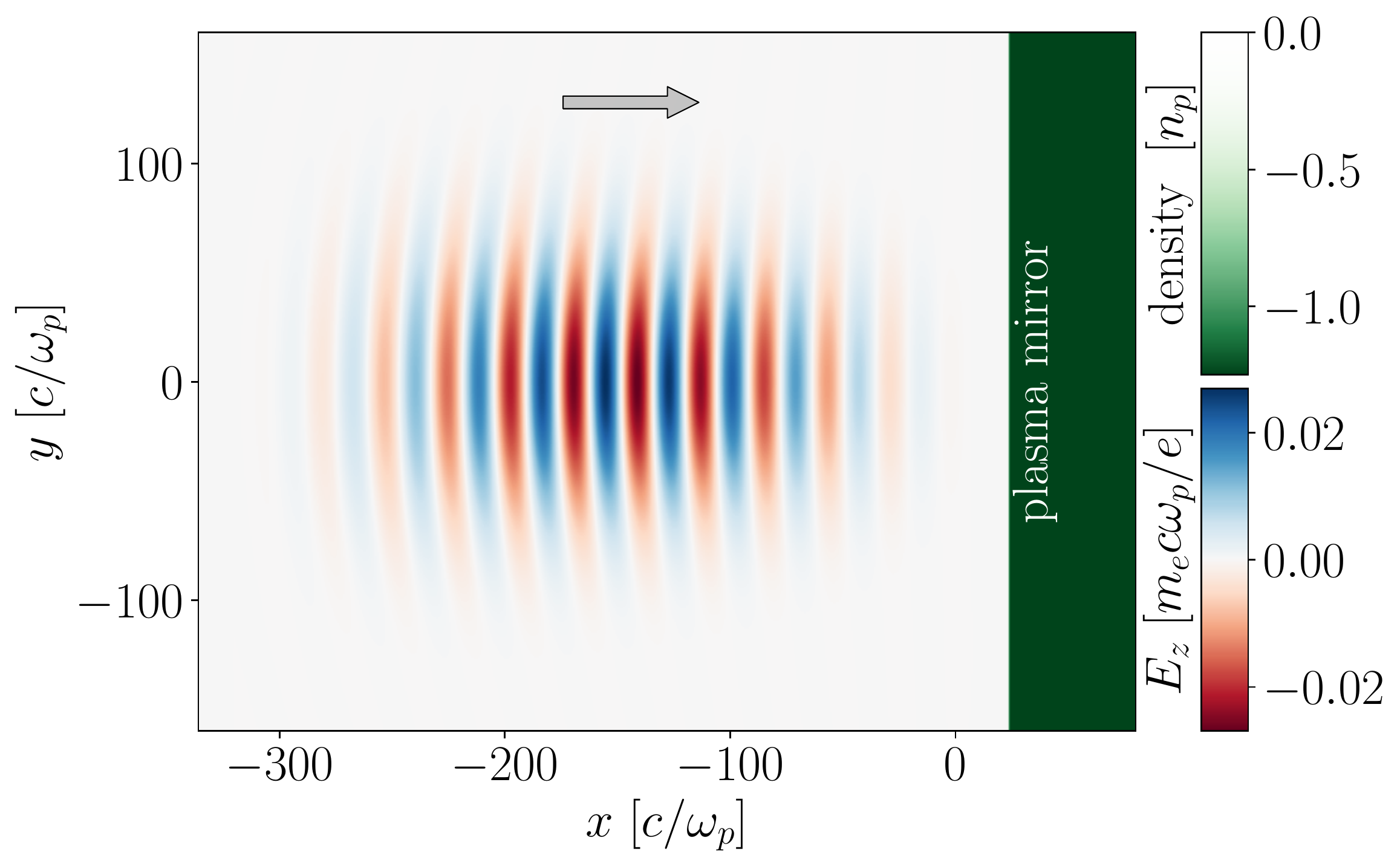}
  \caption{Reflection radiation simulation setup. A tightly focused gaussian laser pulse propagates from left to
   right towards an overdense plasma target.}
  \label{fig:hhg_setup}
\end{figure}

In this section, we present 3D Osiris simulations
of an HHG scenario where the laser propagates in the longitudinal $x$ direction and is linearly polarized
along the transverse $z$ direction. The laser uses a $\sin^2$ temporal profile with 12 full periods
($T_0=8\pi~\omega_p^{-1}$, $\omega_0=\omega_p/4$) and Gaussian perpendicular profile with spot-size 
($W_0=2\lambda_0$, $\lambda_0$ is the central laser wavelength). The plasma mirror consists in an overdense 
plasma slab with plasma frequency $\omega_p$ (and density $n_p$, $16$ times larger than the critical density $n_c$ for that laser pulse)
with thickness 100~$c/\omega_p$, much higher 
than the non-relativistic plasma skin-depth
($l_s\sim~c/\omega_p$ in this case). As the laser gets reflected, we capture the reflected fields both
in the PIC grid through Maxwell's equations and in a virtual detector through RaDiO. We chose to compute
the radiation emitted by all plasma electrons located within the plasma cylinder with a radius of three laser spot
sizes around the focus. The virtual cartesian detector was located at $x=-160~c/\omega_p$, ranging
from $y=-160~c/\omega_p$ to $160~c/\omega_p$, with temporal resolution $dt_{\text{det}}=0.0384~\omega_p^{-1}$, 
about five times smaller than the PIC temporal resolution $dt_{\text{PIC}}=0.1792~\omega_p^{-1}$.  The PIC simulation box
ranged from $x=-288~c/\omega_p$ to $x=108~c/\omega_p$, with a resolution $dx=0.96~c/\omega_p$ in the longitudinal direction and from
$y,z=-160~c/\omega_p$ to $y,z=160~c/\omega_p$ with resolution $dy,dz=0.32~c/\omega_p$ in the transverse direction. 
This PIC grid is able to resolve 26 points per laser wavelength. Each cell contains 16 simulation particles.
A 2-D slice of the setup is shown in \figref{fig:hhg_setup}, the laser propagates from left to right.

\begin{figure}[H]
  \centering
  \begin{tikzpicture}
    \node at (0,0) { \includegraphics[width=0.7\textwidth]{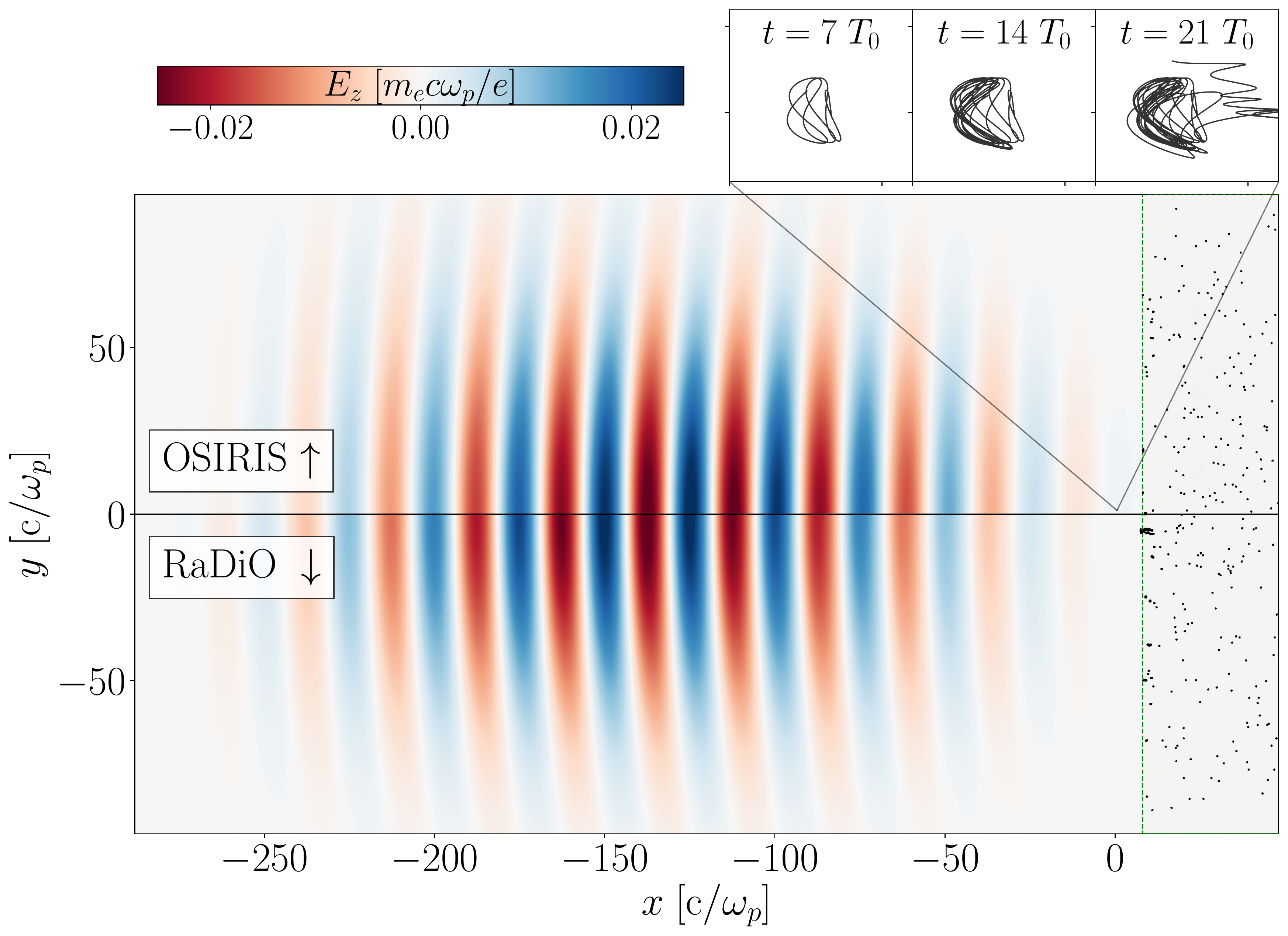}    };
    \node at (-3.8,2.4) {a)};
    \end{tikzpicture}
          \begin{tikzpicture}
    \node at (0,0) {  \includegraphics[width=0.7\textwidth]{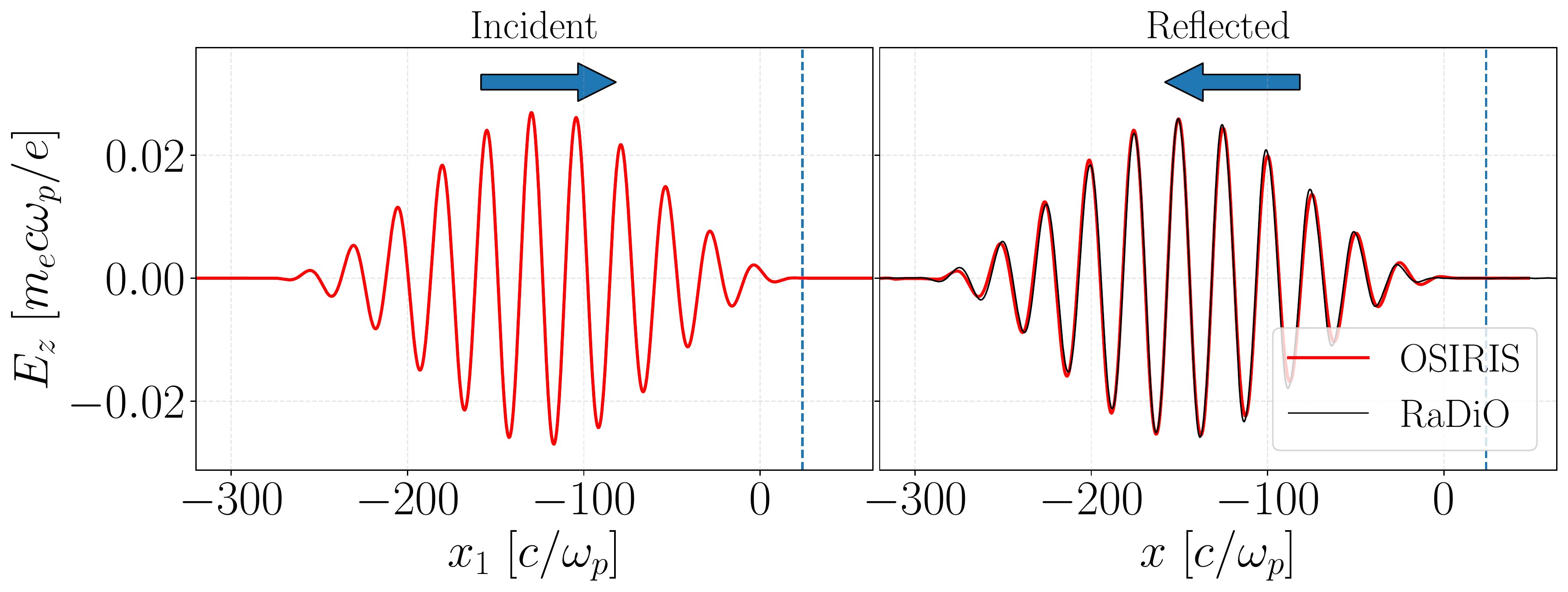}    };
    \node at (-3.8,1.5) {b)};
    \node at (0.5,1.5) {c)};
    \end{tikzpicture}    
  \caption{Trajectories of a random sample of 512 plasma particles under the influence of a low intensity laser ($a_0=0.1$).
   The zoomed-in region shows a particle performing the figure-8 motion induced by the incident laser (top).
    Comparison between the reflected laser profile given by the PIC grid (upper half) and by RaDiO  (lower half) at $x=-160~c/\omega_p$.
     Comparison between incident  and reflected beams as captured by the standard PIC algorithm and by RaDiO (bottom). 
     The laser pulses are properly described in both situations with more than 20 points per wavelength.}
  \label{fig:low_ao_track}
\end{figure}
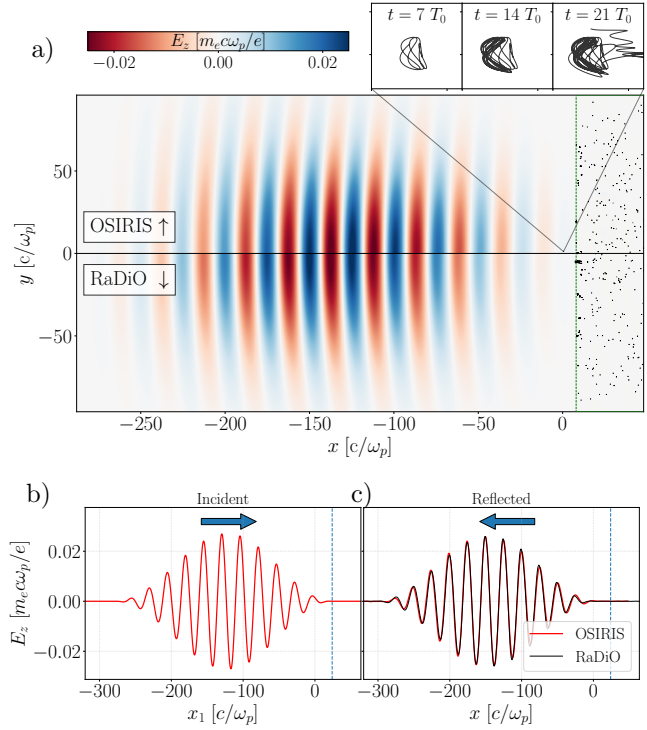

We start by capturing the radiation in the absence of HHG, by using a non-relativistic laser intensity,
 with peak normalized vector potential $a_0=0.1$. \figref{fig:low_ao_track}, top, shows the trajectories
  of a random sample of 512 plasma particles. The zoomed-in region clearly displays the typical
\textit{figure-8}-like motion induced in the plasma particles by the laser pulse. This motion originates the radiation,
 which is captured both in the PIC grid and in the virtual radiation detector. By comparing the radiation in the
  detector to the reflected pulse in the PIC grid~(\figref{fig:low_ao_track}, bottom), we show that the beam reflection
   is a direct result of the charged particles' trajectories induced by the incident beam.

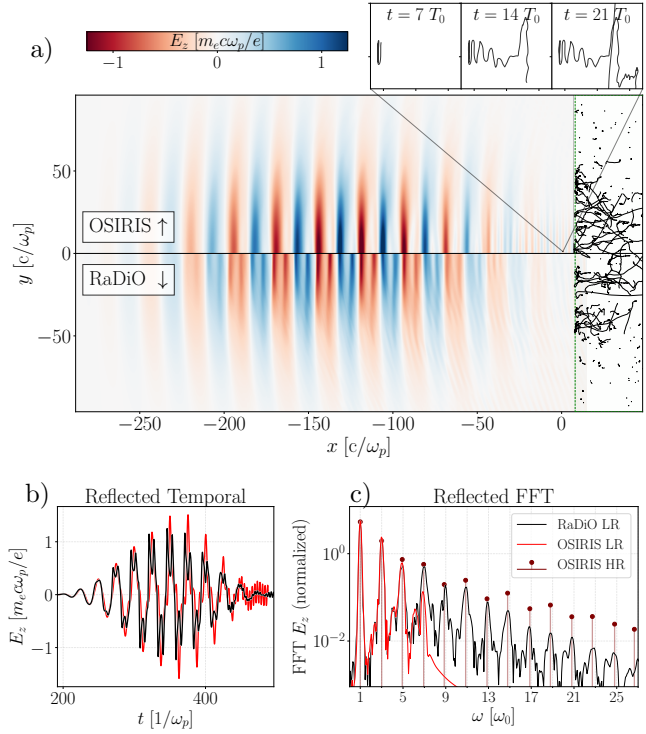
\begin{figure}[H]
  \centering
  \begin{tikzpicture}
    \node at (0,0) { \includegraphics[width=0.7\textwidth]{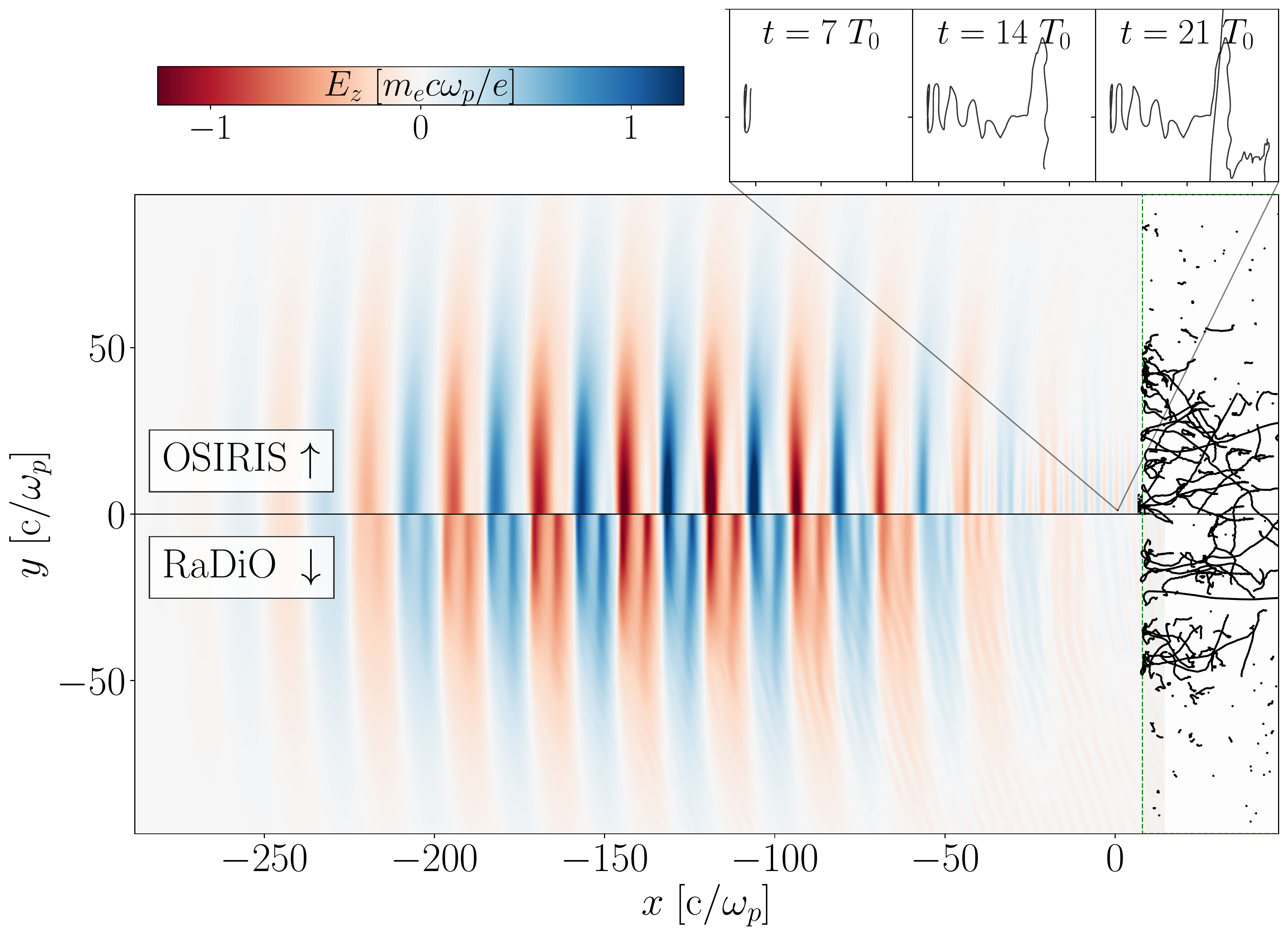}    };
    \node at (-3.8,2.4) {a)};
    \end{tikzpicture}
          \begin{tikzpicture}
    \node at (0,0) {  \includegraphics[width=0.7\textwidth]{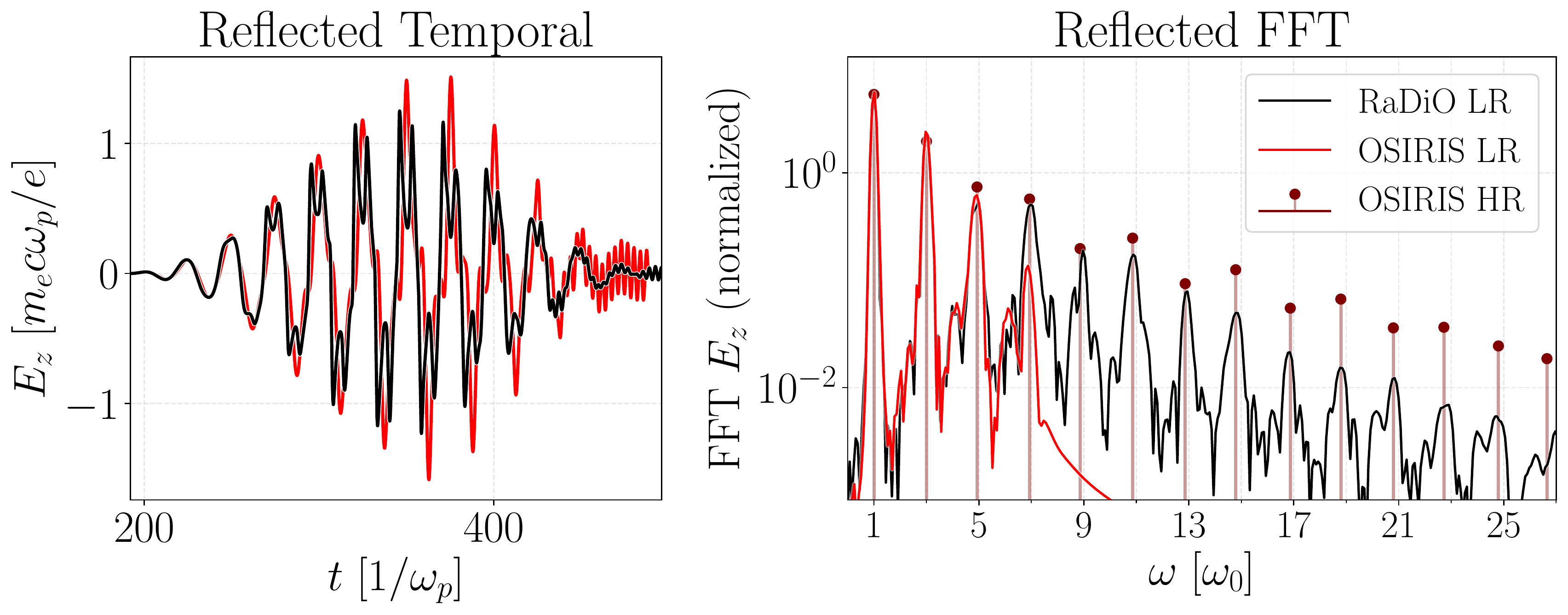}    };
    \node at (-3.8,1.5) {b)};
    \node at (0.5,1.5) {c)};
    \end{tikzpicture}    
  \caption{Trajectories of a random sample of 512 plasma particles (a) after the reflection a low intensity laser ($a_0=4.2$). 
  The zoomed-in region shows a particle performing the figure-8 motion induced by the incident laser.
  Comparison between the reflected laser profile given by the PIC grid (upper half) and by RaDiO  (lower half)  $x=-160~c/\omega_p$.
  Spatiotemporal (b) and frequency spectrum (c) of the reflected high intensity laser beam. } 
  \label{fig:high_a0_track}
\end{figure}

Next to investigate a scenario with strong HHG, we used a high-intensity laser ($a_0=4.2$) in a setup
similar to the one shown on~\figref{fig:hhg_setup}. In this case, we see the clear effect of the
increased intensity on the trajectory of the sampled particles~(\figref{fig:high_a0_track}, top),
with a similar figure-8 motion  for the first few laser periods, but with increased amplitude 
overall and stronger deviation from the standard figure-8 motion.

As a result of this more extreme 
motion, the reflected laser beam is noticeably different from the incident beam. This
is made clear in the comparison shown at the bottom of \figref{fig:high_a0_track}. The
differences between incoming and reflected laser pulse electric field profile are due
to the existence of high laser harmonics present in the reflected beam. The presence 
of the high harmonics is also clearly visible in the spectrum of \figref{fig:high_a0_track}.
The frequency spectrum shows that the reflected laser captured by RaDiO contains at least 13 
harmonics, while the PIC algorithm, which only resolves the plasma relevant scales correctly
captures the emission of the first 4 odd harmonics. The PIC grid is able to resolve the original
harmonic with 26 points per wavelength, but as the harmonic order increases, 
past the 7th order only RaDiO's resolution can capture the signal correctly. In this case the RaDiO
frequency spectrum captures frequencies at least 4 times higher than the OSIRIS PIC grid, being able
to capture harmonics at least until the 25th order, as expected from the employed laser intensity ($a_0=4.2$).

\section{\label{sec:concl} Conclusions}

The radiation diagnostic for OSIRIS (RaDiO) was successfully implemented, benchmarked and tested
in several scenarios, including production runs.
While not described here, it should be also be noted that the algorithm was  
fully parallelized allowing for large  simulations. RaDiO is a novel radiation 
diagnostic that captures the spatiotemporal features of high frequency radiation in PIC codes.
A key aspect of our algorithm is the development of a temporal interpolation scheme for 
depositing radiation. This is essential to preserve the continuous character of radiation 
emission and to obtain correct values for the amplitude of the radiated fields.
The algorithm is general and only requires knowledge about the trajectories of a an
arbitrarily large ensemble of charged particles ($>10^6$) thus we can apply it to 
generally enhance the capabilities of any algorithm that predicts the trajectories of 
charged particles, apart from PIC codes.
We described the implementation of RaDiO into OSIRIS and provided benchmarks with well 
established theoretical models for synchrotron emission. These comparisons showed excellent 
agreement, therefore adding a high level of confidence to future runs.

We also provided an illustration where we used RaDiO to probe the spatiotemporal features 
of radiation emitted in the context of laser reflection by a plasma mirror. At lower laser 
intensities, RaDiO fully recovers the PIC simulation result. This further confirms the 
validity of RaDiO in a setting where temporal and spatial coherence effects are critical. 
A simulation at higher laser intensity demonstrated the generation of high harmonics beyond 
the predictions of the PIC algorithm, showing that RaDiO allows for a complete 
characterization of the reflected beam along with all the harmonics, without 
increasing the overall PIC resolution, and effectively demonstrating that RaDiO can 
be effectively used to predict high frequency radiation from PIC 
codes~\cite{10.1038/s41567-020-0995-5}.

RaDiO is a flexible diagnostic tool that can can be further expanded to include additional 
features such as higher order interpolation schemes, for example using an advanced 
particle pusher recently developed~\cite{LI2021110367}, the option to compute 
the electromagnetic field potentials in addition to the electromagnetic fields, 
or the capability to convert radiation to/from relativistic Lorentz boosted frames. 
Although this diagnostic does not interact with the particles, it could also be
employed together with a QED code that captures radiation
reaction and affects the particle's trajectories and capture radiation
compatible with QED effects as long as the emission is purely classical.
Because it captures the radiation in space and in time, RaDiO may also be useful in 
describing the production of spatiotemporally structured beams~\cite{PhysRevLett.113.153901}.



\section*{Acknowledgments}

We acknowledge the Partnership for Advanced Computing in Europe (PRACE) for access to the Leibniz Research Center on SuperMUC and the Barcelona Supercomputing Center on Marenostrum 4. This work was partially supported  by the  EU Accelerator Research for Innovation for European Science and Society (EU ARIES) under  grant agreement no. 738071 (H2020-INFRAIA-2016-1). 
JV acknowledges the support of FCT (Portugal) Grant No. IF/01635/2015/CP1322/CT0001 and MP acknowledges the support of FCT (Portugal) Grant No. PD/BD/150411/2019.

\bibliography{library}

\end{document}